\documentclass[review]{elsarticle}

\journal{Journal of \LaTeX\ Templates}

\biboptions{sort&compress}

\begin{document}

\newcommand{\didv}{$\mathrm{d}I/\mathrm{d}V$\,}

\begin{frontmatter}

\title{Stable $\pi$-radical 2,2-diphenyl-1-picrylhydrazyl (DPPH) adsorbed at the elbows of $22\times\sqrt{3}$ reconstructed Au(111)}

\author[jku]{Radovan Vranik} 

\author[jku]{Vitalii Stetsovych} 

\author[jku]{Simon Feigl} 

\author[jku,lit]{Stefan M\"{u}llegger\corref{correspondingauthor}}

\cortext[correspondingauthor]{Corresponding author. Email address: stefan.muellegger@jku.at}

\address[jku]{Institute of Semiconductor and Solid State Physics,  Johannes Kepler University Linz, Altenberger Strasse 69, 4040 Linz, Austria.} 
\address[lit]{Linz Institute of Technology,  Johannes Kepler University Linz, Altenberger Strasse 69, 4040 Linz, Austria.}

\begin{abstract}
Stable organic radicals serve as model systems for investigating metal-free magnetic phenomena at (sub)nanometer length scales. 
We have investigated at the single-molecule level the stable $\pi$-radical 2,2-diphenyl-1-picrylhydrazyl (DPPH) adsorbed at the elbow sites of the $22\times\sqrt{3}$ reconstructed Au(111) surface. 
Low-temperature scanning tunneling microscopy (STM) and -spectroscopy under ultrahigh vacuum conditions at 8~K reveal structural as well as frontier-orbital related electronic details of DPPH/Au(111). 
A Kondo-like spectroscopic signature indicates preservation of the unpaired electron spin state. 
\end{abstract}


\end{frontmatter}

\section{Introduction}
\begin{figure}
\includegraphics{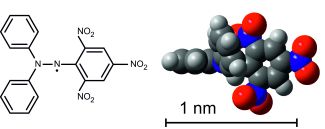}
\caption{2,2-Diphenyl-1-picrylhydrazyl (DPPH) molecule. Left: Chemical structure. Right: Molecular structure obtained by geometry optimization from first-principles in gas phase.}
\label{fig:schema}
\end{figure}

Stable molecular radicals adsorbed on metal surfaces serve as model systems in diverse research fields such as molecule-based spinterfaces \cite{Ratera2012,Requist2014,Mugnaini2015,Wu2015,Bocquet2019,Casu2018}, molecular spintronics \cite{Torrent2009,MasTorrent2012,Requist2016,Karan2016a,Karan2016b,Pavlicek2017,Low2019}, and heterogeneous catalysis \cite{Mullegger2017c}. 
The molecule 2,2-diphenyl-1-picrylhydrazyl (DPPH) shown in Fig.~\ref{fig:schema} is an archetypal stable  $\pi$-radical well-known in the literature, see for instance the review by Foti et al. \cite{Foti2015}. 
The highest occupied molecular orbital (HOMO) of the DPPH molecule contains a single electron, resulting in an unpaired electron spin configuration. 
Due to the highly isotropic electronic $g$-factor of 2.0037 \cite{Weil2007}, DPPH has been commonly applied as calibration standard for electron paramagnetic resonance experiments as well as diagnostic reagent for studying chemical kinetics \cite{Weil2007,Foti2008}. 
Clusters of DPPH molecules have been successfully applied for detecting spin noise signals close to the single-spin limit \cite{Mannini2007a,Messina2007}. 

Combined experimental and theoretical studies of DPPH in different solvents have reported that the spin density of the unpaired electron is mainly shared between the two central nitrogen atoms, see Fig.~\ref{fig:schema}, and is only slightly delocalized into the aromatic rings \cite{Dalal1973,Dalal1982,Foti2008}. 
The molecular structure of DPPH has been described as nearly planar in different liquid solvents \cite{Dalal1973,Dalal1982,Messina2009b}, whereas in frozen solution the molecular structure appears kinked and the picryl ring twisted relative to the N-N-C plane \cite{Messina2009b}.
To date, the majority of experiments on DPPH has focused on bulk phase or solution \cite{Foti2015} and studies at the single-molecule level remain scarce. 

Here we report single-molecule experiments on individual DPPH molecules by low-temperature scanning tunneling microscopy (STM) and -spectroscopy (STS). 
We have studied isolated DPPH molecules physisorbed on a Au(111) single crystal surface under ultrahigh vacuum conditions at 8~K. 
The presence of a Kondo-like electric conductance resonance indicates that DPPH preserves the unpaired electron spin upon adsorption on Au(111). 

\section{Materials and methods} 
In-situ preparation of our samples was performed in ultrahigh vacuum (UHV) at a base pressure of $< 2 \cdot10^{-9}$~mbar. 
Experiments were carried out on a commercial CreaTec low-temperature STM operated at $< 1 \cdot10^{-10}$~mbar and 8~K. 
The substrate was a Au(111) single crystal purchased from Surface Preparation Laboratory. 
The Au(111) surface was cleaned in UHV by repeated cycles of sputtering with Ar$^+$ ions with an energy of 600~eV followed by thermal annealing at 720~K. 
DPPH (C$_{18}$H$_{12}$N$_{5}$O$_{6}$, molecular weight 394.32~amu, CAS number 1898-66-4) was purchased from Sigma Aldrich and  degassed up to 120$^\circ$C for $>8$~hours at UHV conditions. 
The DPPH molecules were thermally evaporated from a quartz crucible heated to 393~K onto the Au(111) surface held at room temperature, while maintaining a base pressure of $<2 \cdot 10^{-9}$~mbar. 
We have determined the coverage of DPPH molecules on the sample with the help of the topographic STM images as shown exemplarily in Fig.~\ref{fig:overview}. 
The single DPPH molecule adsorbed on Au(111) covers a surface area of circa 0.9~nm$^2$. 

The STM tip was prepared from a polycrystalline tungsten wire by electrochemical etching and subsequent deoxidizing in UHV by heating above 1070~K. 
The apex of the STM tip was routinely coated with gold atoms by controlled in-situ indentation into the substrate by typically $<1$~nm at 8~K. 
Prior to each STS measurement the state of the STM tip was carefully checked by reproducing the well-known step-like signature of the Au(111) surface state \cite{Berndt1999}. 
The \didv signal was obtained with lock-in technique and a sinusoidal modulation of the sample bias voltage with an amplitude of 12~mV zero-to-peak at a frequency of 0.773 to 5~kHz. 
Spectroscopic images were recorded at constant-current imaging conditions. 

\section{Results and discussion}
\subsection{Adsorption Configuration}
\begin{figure}
\includegraphics{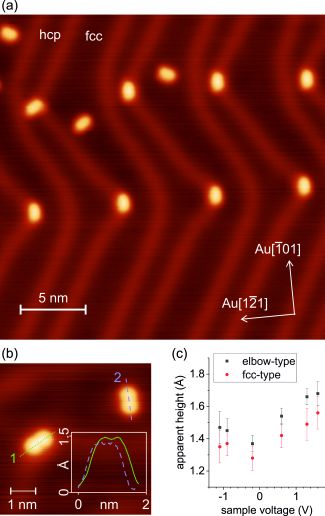}
\caption{(a) STM image of $<0.02$~monolayers DPPH on Au(111), $25\times25$~nm$^2$, $-0.9$~V, $55$~pA, $z$-scale: 200~pm. (b) STM image of isolated DPPH molecules of elbow-type (1) and fcc-type (2), $6\times6$~nm$^2$, $+0.02$~V, $100$~pA, $z$-scale: 160~pm; inset: height profiles of elbow- and fcc-type molecules recorded along the lines 1 (green) and 2 (blue), respectively. (c) Bias-dependence of the apparent height of elbow- and fcc-type DPPH molecules in STM topographic images.}
\label{fig:overview}
\end{figure}

The Au(111) single crystal surface exhibits a characteristic $22\times\sqrt{3}$ surface reconstruction \cite{Barth1990,Chambliss1991,Burgi2002} that forms a regularly patterned array of under-coordinated Au surface atoms (elbow sites) at which molecules can selectively adsorb \cite{Yokoyama2001b,Mullegger2011b}. 
We utilize this regular nano-array for obtaining spatially isolated DPPH molecules on the Au(111) surface.  
Figure~\ref{fig:overview}a shows a representative STM image of the sample surface after depositing $<0.02$~monolayers of DPPH at room temperature and subsequent cooling to 8~K. 
The individual DPPH molecules are imaged by STM as oval-like protrusions. 
DPPH decorating the elbow sites of Au(111) is denoted as elbow-type in the following. 
Notice, that for various other hydrocarbon molecules \cite{Boehringer1999} a similar preferential decoration of the elbow sites of the surface-reconstructed Au(111) has been reported in the literature, including the stable $\pi$-radical 1,3-bis-diphenylene-2-phenylallyl \cite{Mullegger2012b}.
DPPH molecules adsorbed on fcc and hcp regions are herein denoted as fcc- and hcp-type, respectively.  
The hcp-type has been very rarely observed on our samples with the very low coverage, and we therefore neglect it in the rest of this work. 

Figure~\ref{fig:overview}b shows elbow- and fcc-type DPPH molecules at increased magnification and with tunnel parameters ($+20$~mV, 100~pA) optimized for imaging topographic details: 
Both elbow- and fcc-type molecules have a dumbbell-like contour. 
The lateral size of single DPPH molecules has been determined from height profiles of our topographic STM images as illustrated exemplarily in the inset of Fig.~\ref{fig:overview}b. 
For the single DPPH molecule on Au(111) imaged at $+20$~mV we obtain a length of $1.20\pm0.05$~nm and a width of $0.75\pm0.05$~nm. 
The size is bias dependent and is found to increase slightly at other voltages. 
For instance, at a bias voltage of $-180$~mV ($-900$~mV) the length and width are 1.23~nm (1.39~nm) and 0.80~nm (1.03~nm), respectively. 
These values are in good agreement with those reported by Messina et al. \cite{Messina2009b} for DPPH imaged by STM at ambient conditions and similar to the dimensions of the DPPH molecule in gas phase (Fig.~\ref{fig:schema}), as well. 
Fig.~\ref{fig:overview}c shows the topographic height of fcc- and elbow-type DPPH molecules determined from height profiles of the STM images at different bias voltage values. 
All height values were taken relative to the fcc region of the reconstructed Au(111) surface. 
Accordingly, the height of the elbow-type molecules in Fig.~\ref{fig:overview}c appears systematically larger by ca. 0.15~\AA ngstrom because of the vertical corrugation of the elbow sites relative to the fcc region \cite{Barth1990}. 
A further discussion of the topographic dimension of the molecule is given in the supplementary information. 

The anisotropic, dumbbell-like shape of the DPPH molecule (Fig.~\ref{fig:overview}b) makes it possible to determine the azimuthal orientation relative to the substrate atomic lattice of the $22\times\sqrt{3}$ reconstructed Au(111) surface. 
The latter is characterized by the high-symmetry directions Au$[\bar{1}01]$ and Au$[1\bar{2}1]$ directions \cite{Chambliss1991} displayed in Fig.~\ref{fig:overview}a. 
Analyzing more than 200 individual DPPH molecules in our STM images, we have found that the elbow-type molecules adsorb predominantly with their long semi-axis aligned tangentially to the apex of the elbow site, i.e. parallel to the Au$[\bar{1}01]$ direction. 
The fcc-type molecules prefer any of the six symmetry-equivalent Au$\left\langle1\bar{1}0 \right\rangle$ directions. 

\begin{figure}
\includegraphics[width=7.5cm]{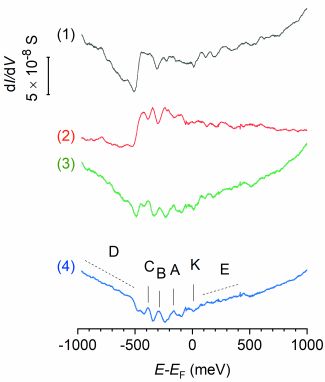}
\caption{Tunnel conductance (\didv) point spectra. All curves are vertically separated for better clarity. Curve~1: Single elbow-type DPPH molecule on Au(111); STM tip was positioned close to the rim of the molecule. Curve~2: Pristine Au(111) surface. Curve~3: Difference spectrum obtained by numerically subtracting curve~2 from curve~1. Curve~4: Average difference spectrum of a DPPH molecule on Au(111) obtained by averaging 30 difference spectra that have been individually recorded on the same DPPH molecule.}
\label{fig:spec}
\end{figure} 

\subsection{Electronic Properties}
We have investigated the frontier-orbital related electronic properties of single DPPH molecules on Au(111) with the help of \didv point spectroscopy as well as spectroscopic imaging \cite{Chen2008}. 
Herein we have focused on elbow-type molecules, which have  several advantages for performing STS experiments as compared to fcc and hcp types: 
(i) At such ultra-low DPPH coverages applied herein, elbow-type molecules are by far the most abundant species, which enables to study a sufficiently large number of different individual DPPH molecules for obtaining reliable spectroscopic information. 
(ii) Elbow-type molecules have a highly reproducible adsorption position and azimuthal orientation, thus minimizing undesirable geometrical effects during spectroscopy. 
(iii) Elbow-type molecules are stronger bonded and thus more stable during the STS experiment, improving the quality and reproducibility of the results. 
All these benefits have turned out to be particularly important for performing STS experiments on DPPH/Au(111) at the single-molecule level. 

Fig.~\ref{fig:spec} curve~1 shows a representative single \didv point spectrum of elbow-type DPPH molecules on Au(111). 
The most prominent spectroscopic feature of curve~1 is a step close to $-470$~meV. 
The spectrum of pristine Au(111), curve~2, exhibits a similar step, which is well-known to originate from the onset of the (two-dimensional) electronic surface state of Au(111) \cite{Burgi2002}. 
The appearance of similar steps in, both, curve~1 and 2 indicates that the electronic surface state of Au(111) is preserved upon the adsorption of the DPPH molecule and points to a rather weak electronic interaction between DPPH and the gold substrate \cite{Nicoara2006,Koslowski2011a,Koslowski2011b}. 
Apart from the step, curve~1 lacks any additional pronounced \didv features and appears rather similar to curve~2 throughout the energy range of $\pm1000$~meV. 
In order to distinguish in the experimental spectra the electronic contributions of substrate and DPPH, we subtract curve~2 from curve~1, resulting in the difference spectrum, curve~3. 
Notice, that such background subtraction procedure is appropriate in the present case, because of the weak molecule-substrate electronic interaction, as recently pointed out by Wahl et al. \cite{Wahl2008}. 
Since the remaining \didv features after background subtraction are tiny, we have averaged several independent difference spectra. 
The averaging improves the reliability of the observed spectral features by increasing the signal-to-noise ratio. 
In addition, undesirable effects of the STM tip and its lateral position relative to the DPPH molecule are averaged out. 

Fig.~\ref{fig:spec} curve~4 shows the average of 30 independent difference spectra that have been recorded with three different STM tips and at ten different lateral positions of the STM tip across the DPPH molecule. 
A close inspection of curve~4 reveals distinct \didv peaks at energies of $-160$, $-300$, and $-390$~meV, labeled A--C, band-like features, D and E, as well as a tiny feature close to zero bias, labeled K. 
All features A--K have survived the averaging procedure and are clearly discernible in curve~4 -- while being absent in the spectrum of the substrate (compare curves~2 and 4). 
Accordingly, we attribute features A--K to the involvement of electronic contributions from the molecular frontier-orbitals of the adsorbed DPPH molecule. 
Peaks A--C have a full-width at half maximum of circa 55~meV. 
This value is rather small compared to typical values of the full-width at half maximum of about 100 to 300~meV previously reported for other radical molecules on Au(111) \cite{Iancu2006a,Gao2007,Zhao2008,Mugarza2011,Mullegger2012b,Iancu2014}. 
The small width of the electronic states A--C indicates that electronic contributions from the gold substrate are weak \cite{Mullegger2011b}. 
In contrast, feature D is exceptionally broad. 
It appears below $-550$~meV and extends to beyond $-1000$~meV. 
Its almost band-like appearance indicates increased electronic contribution of substrate electronic states. 
The empty-states regime of curve~4 (at positive energies) appears almost structureless. 
The only discernible feature is a tiny band-like feature, labeled~E, that extends from closely below $+100$~meV to approximately $+450$~meV. 
The empty-states regime exhibits no discernible peak-like features. 
Further details are given in the final discussion below.

\begin{figure}
\includegraphics{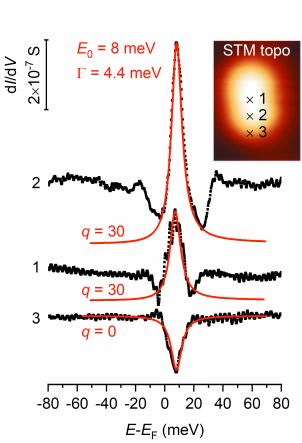}
\caption{Experimental tunneling conductance spectra, curves~1--3, recorded with the STM tip at positions 1--3. Red lines are numerical fits by Fano profiles with fitting parameters $E_0$, $\Gamma$ and $q$ (see text for details). Inset: STM topographic image of a single DPPH molecule on Au(111), $1.8\times3.5$~nm$^2$, $-180$~mV, 0.86~nA, $z$-scale 160 pm. Crosses~($\times$) labeled~1--3 mark the different lateral positions of the STM tip during point spectroscopy.}
\label{fig:kondo}
\end{figure}

Figure~\ref{fig:kondo} shows \didv spectra of feature~K with increased detail, revealing a characteristic peak- or dip-like shape depending on the lateral position of the STM tip across the DPPH molecule. 
Labels 1--3 indicate the different STM tip positions during spectroscopy, resulting in the respective \didv curves~1--3. 
Notice, that all displayed spectra of Fig.~\ref{fig:kondo} have been corrected for the effects of voltage modulation during lock-in detection of the \didv signal \cite{Klein1973,Li1998} and finite temperature \cite{Klein1973}. 
The shape of feature~K changes between peak and dip. 
When the STM tip is over the molecule, positions~1 and 2, it takes the form of a peak and close to the rim, position~3, it takes the form of a dip. 
The red curves in Fig.~\ref{fig:kondo} represent numerical fits of the experimental data with a Fano profile \cite{Fano1961,Plihal2001}. 
The latter can be expressed as 
\begin{equation}
\frac{\mathrm{d}I}{\mathrm{d}V}(\epsilon)\propto\frac{(\epsilon+q)^2}{\epsilon^2+1}	
\end{equation}
where $\epsilon:=(eV-E_0)/\Gamma$ and $V$ denotes the sample bias voltage. 
In each of the curves~1--3, we have obtained least-squares fits of the peak (dip) with a coefficient of determination of $R^2>0.9$, using the same numerical values for the center position, $E_0 =+8$~meV, and for the half width at half maximum, $\Gamma=4.4$~meV. 
Based on the good agreement in Fig.~\ref{fig:kondo} between the Fano fits and the experimental data, we may attribute  feature~K to a Kondo signature \cite{Madhavan1998,Li1998b,Mullegger2013a}. 
Accordingly, the Kondo temperature of elbow-type DPPH/Au(111) has an approximate value of $\Gamma/k_\mathrm{B}=51$~K. 

The different shape of either a peak or a dip is determined by the form factor $q$. 
We have obtained values of $q=30$ for curves~1\&2 and $q=0$ for curve~3, respectively. 
These values reflect the symmetrical course taken by, both, peak and dip in Fig.~\ref{fig:kondo}. 
The form factor is proportional to the ratio of electrons tunneling to the Kondo resonance and to the continuum of substrate states, respectively \cite{Plihal2001,Ternes2009}. 
Thus, curves~1 and 2 indicate that electrons tunnel predominantly between the STM tip and the Kondo resonance, while curve~3 points to predominant tunneling between the STM tip and the continuum of substrate states.

\section{Final discussion}
The observed Kondo-like resonance of elbow-type DPPH (Fig.~\ref{fig:kondo}) indicates that the unpaired electron spin configuration is preserved upon adsorption on Au(111) at UHV conditions. 
For fcc-type DPPH, point spectroscopy of single individual molecules is complicated by surface diffusion, and we have obtained strong indications for a Kondo-like behavior, as well, see supplementary information. 
In this respect, DPPH behaves similar to other stable pi radicals like the Koelsch radical \cite{Mullegger2012b} or the Blatter radical \cite{Patera2019}. 

All Kondo resonances shown in Fig.~\ref{fig:kondo} exhibit very similar center positions ($E_0$ values). Following the argumentation of Knaak et al. \cite{Knaak2017}, the independence of $E_0$ from the STM tip position across the molecule indicates that one single orbital of the radical molecule contributes to the Kondo screening. 

The center of the Kondo-like resonance lies slightly above the Fermi level in the empty-states regime (Fig.~\ref{fig:kondo}). 
This finding, which is here expressed by the positive value of $E_0$, has an important implication when considering Kondo behavior in the framework of Fermi liquid theory \cite{Hewson1993}: 
The time-average electronic occupation of the SOMO is less than one. 
Such partial depopulation of the SOMO may originate from electronic charge transfer between the radical molecule and the substrate \cite{Krull2013}, causing electric charge localization at the adsorption site of the molecule. 
Localized electric charge at the surface is well known to enhance the scattering of electronic states, which can be observed by STM spectroscopic imaging \cite{Repp2004}. 

\begin{figure}
\includegraphics{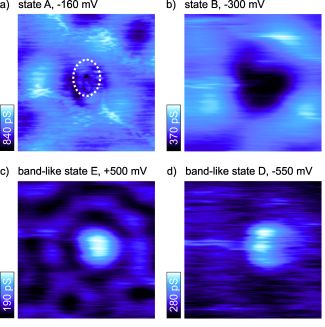}
\caption{Spectroscopic images ($5\times5$~nm$^2$) of a single elbow-type DPPH molecule on Au(111)   recorded at constant tunnel current (860, 650, 650, 650 pA) and for different sample voltages. 
The dotted-line oval marks the topographic outline of the DPPH molecule as guide to the eye.}
\label{fig:map}
\end{figure}

Figure~\ref{fig:map} shows spectroscopic images of a single elbow-type DPPH molecule on Au(111) at different sample bias voltages. 
Notice, that each image has been scaled to an individual range of conductance (color scale) in order to achieve maximum image contrast. 
As a guide to the eye, the dotted-line oval marks the approximate topographic circumference and position of the DPPH molecule determined from STM topographic imaging over the same image frame. 
Figures~\ref{fig:map}a--c exhibit standing-wave patterns due to the scattering of the surface-state electrons at the adsorbed DPPH molecule. 
The observed patterns appear rather similar to those observed commonly on pristine Au(111) due to the well-known scattering at step edges and adatoms \cite{Hasegawa1993}.   
Thus, Figs.~\ref{fig:map}a--c provide no conclusive evidence for a significantly increased charge localization near the DPPH molecule, suggesting that a possible charge transfer between the molecule and the substrate is rather small -- similar to the case of BDPA/Au(111) \cite{Mullegger2012b}. 
As expected, no standing wave pattern is observed in the spectroscopic image of state~D (Fig.~\ref{fig:map}d) because the energy lies below that of the surface state.  

An additional implication of the positive value of $E_0$, observed here for DPPH, is that the alignment of SOMO and SUMO relative to the Fermi level is expected to become asymmetric, with SOMO lying closer in energy to the Fermi level and SUMO lying farther away \cite{Hewson1993}. 
Based solely on the spectroscopic data of DPPH/Au(111), the unambiguous determination of SOMO and SUMO energies is difficult without the additional support of first principles calculations. 
The highest occupied electronic state clearly discernible by point spectroscopy is peak~A at $-160$~meV, see Fig.~\ref{fig:spec} curve~4. 
However, the straightforward attribution of peak~A as SOMO is not possible, because the existence of other resonances, lying even closer in energy to the Fermi level cannot be savely ruled out \cite{Villagomez2009,Amokrane2017}. 
Similar arguments complicate the attribution of the SUMO. 
The empty states regime of Fig.~\ref{fig:spec} curve~4 exhibits only very tiny \didv signatures, similar to the case of the Blatter radical studied recently \cite{Patera2019}. 
Nevertheless, curve~4 provides conclusive evidence for the existence of molecule-related electronic states within a few hundred milli-electronvolts close to the Fermi level in, both, the filled- and empty states regimes. 
The respective energy separation between the highest occupied state and the lowest unoccupied state is less than 1~eV, see curve~4, which is much smaller than the energy gap of about $2.4$~eV reported for DPPH by optical absorption spectroscopy in apolar solvents \cite{Foti2008}. 

\section{Conclusions}
We demonstrate that the stable pi-radical DPPH adsorbed at the elbow sites of the $22\times\sqrt{3}$ reconstructed Au(111) surface preserves the unpaired electron spin. 
We provide tunneling spectroscopy data at the single molecule level, revealing molecule-related electronic states lying only within a few hundred milli-electronvolts close the Fermi level. 
The experimental data presented herein may serve as basis for future theoretical approaches to achieve a first-principles predictive capability of Kondo conductance anomalies across molecular radicals \cite{Requist2014}. 

\section{Acknowledgements}
The authors acknowledge the financial support by the European Research Council (ERC) under the European Union's Horizon 2020 research and innovation programme (grant agreement No 771193) as well as the Government of the Province of Upper Austria together with the Johannes Kepler University Linz (LIT project 2016-1-ADV-002). 

Declarations of interest: none.


\end{document}